\input{aipcheck}

\documentclass[
    ,final
    ]
  {aipproc}

\layoutstyle{8x11single}

\usepackage{float,multicol,subfigure,amsmath, amsthm, amssymb}
\usepackage{palatino}
\usepackage{color}
\usepackage{framed}

\def\rms{r_{\rm ms}}
\def\xms{x_{\rm ms}}
\def\dM{\dot{M}}
\def\dMign{\dot{M}_{\rm ign}}

\def\dMtr{\dot{M}_{\rm trap}}

\def\Kign{K_{\rm ign}}

\def\Ktr{K_{\rm trap}}
\def\dE{\dot{E}_{\nu\bar\nu}}
\def\Teff{T_{\rm eff}}
\def\Lobs{L_{\rm obs}}

\begin{document}

\title{Efficiency of Neutrino Annihilation around Spinning Black Holes}

\classification{ --}
\keywords      {Accretion - accretion disks - gamma ray bursts - Kerr black holes}

\author{I. Zalamea}{
  address={izalamea@phys.columbia.edu}
}

\author{A. M. Beloborodov}{
  address={amb@phys.columbia.edu}
}

\begin{abstract}
A fraction of neutrino emission from GRB accretion disks annihilates
above the disk, creating e+- plasma that can drive GRB explosions.
We calculate the efficiency of this annihilation using the recent
detailed model of hyper-accretion disks around Kerr black holes.
Our calculation is fully relativistic and based on a geodesic-tracing
method. We find that the efficiency is a well-defined function of
(1) accretion rate and (2) spin of the black hole. It is practically
independent of the details of neutrino transport in the opaque zone
of the disk. The results help 
identify the accretion disks whose
neutrino emission can power GRBs.
\end{abstract}

\maketitle

\section{Introduction}
The relativistic model of GRB accretion disks was completed recently
\cite{Chen:2007}. It describes the disk matter down to the last 
stable orbit $\rms$ and gives
energy fluxes carried away by $\nu$ and $\bar{\nu}$ at all radii $r$. 
In the present work, we trace the trajectories of emitted neutrinos, 
and calculate the rate of $\nu\bar{\nu}$ annihilation around the disk. 
This process deposits $e^\pm$ plasma and can play a key role in the formation 
of GRB jets.

Neutrino annhilation was previously calculated in a number of works 
(e.g. \cite{Popham:1999}, 
\cite{Asano:2001},
\cite{Birkl:2007}) 
Our work has two motivations: 
(1) A relativistic calculation has never been done for a realistic accretion 
disk around a spinning black hole. Previous works either used a toy model for 
neutrino source (e.g. an isothermal disk, \cite{Birkl:2007}) or replaced 
neutrino trajectories by straight lines \cite{Popham:1999}. 
(2) The efficiency of $\nu\bar\nu$ annihilation depends strongly on the 
accretion rate $\dot{M}$ and the spin parameter $a$ of the black hole. 
It is desirable to know 
this dependence and identify the range of $\dot{M}$ 
and $a$ where $\nu\bar\nu$ annihilation can provide the observed energy of 
GRB explosions.
\section{Neutrino Emission from the Disk}
Fortunately, 
our final result depends
only on the \emph{energy fluxes} 
 $F_\nu$ and $F_{\bar\nu}$ from the disk surface, which are independent of 
 the neutrino-transport details and are already calculated in 
 \cite{Chen:2007}. The rate of $\nu\bar\nu$ annihilation is 
 insensitive to the exact shapes of $\nu$ and $\bar\nu$ spectra.
 This fact can be demonstrated using two extreme models A and B:

 {\bf Model~A:} Neutrinos $\nu$ and $\bar\nu$ are emitted with the same 
 spectrum as found inside the disk (same temperature $T$ and chemical 
 potential $\mu_\nu$). The spectrum is normalized so that the emerging 
 emission carries away the known energy fluxes $F_{\nu}$ and $F_{\bar\nu}$.

{\bf Model~B:} Neutrinos are emitted with 
the effective surface temperature $T_{\rm eff}$ defined by 
$(7/8)\sigma T_{\rm eff}^4=F_\nu+F_{\bar\nu}$
($\sigma$ is the Stefan-Boltzmann constant). 
The two models give practically the same $\nu\bar\nu$ annihilation rate
(Figure~2). 
Our results confirm the analytical argument that it is sufficient to 
know $\Teff$ to calculate
neutrino annihilation rate \cite{Beloborodov:2008}.
When the disk is efficiently cooled (neutrino losses
almost balance viscous heating), $\Teff$
is given by the standard thin-disk model \cite{Page:1974}:
$\Teff=\Teff^{\rm standard}$.
This model applies to GRB disks in a broad range of accretion rates 
$\dMign<\dM<\dMtr$ \cite{Chen:2007}, where
\begin{equation}
\label{eq:pw}
   \dMign=\Kign\left(\frac{\alpha}{0.1}\right)^{5/3}, \qquad
   \dMtr=\Ktr\left(\frac{\alpha}{0.1}\right)^{1/3}.
\end{equation}
Here $\alpha\sim 0.1$ is the standard viscosity parameter of the 
accretion disk, and the factors $K$ depend on the black hole spin $a$;
e.g. for $a=0.95$ they are $\Kign=0.021$\,M$_\odot$~s$^{-1}$
and $\Ktr=1.8$\,M$_\odot$~s$^{-1}$.
\;\; $\Teff$ of neutrino emission can be approximately described as
   \begin{equation}
   \label{eq:C}
    \Teff(\dM,r)\approx \Teff^{\rm standard}(\dMign,r)\times 
       \left\{ \begin{array}{ll}
      0                 &  \dM<\dMign \\
  (\dM/\dMign)^{1/4}    &  \dMign<\dM<\dMtr \\
  (\dMtr/\dMign)^{1/4}  &  \dM>\dMtr \\
               \end{array}
       \right.
\end{equation}
$\Teff$ is suppressed for $\dM<\dMign$ because the disk is 
not hot enough
to ignite neutrino-emitting reactions. $\Teff$ suddenly increases 
when $\dM=\dMign$ and grows as $\dM^{1/4}$ until $\dM=\dMtr$.
At higher $\dM$, neutrinos become trapped in the disk (advected into 
the black hole) and $\Teff$ saturates. Equation~(\ref{eq:C}) defines
our {\bf Model~C},
which reproduces surprisingly well the more detailed numerical results 
(Figure~2).
This model allows us to obtain an explicit approximate formula for the 
annihilation rate
(eq.~\ref{eq:Edot} below).

\begin{figure}[tp]
\includegraphics[height=0.46\textwidth]{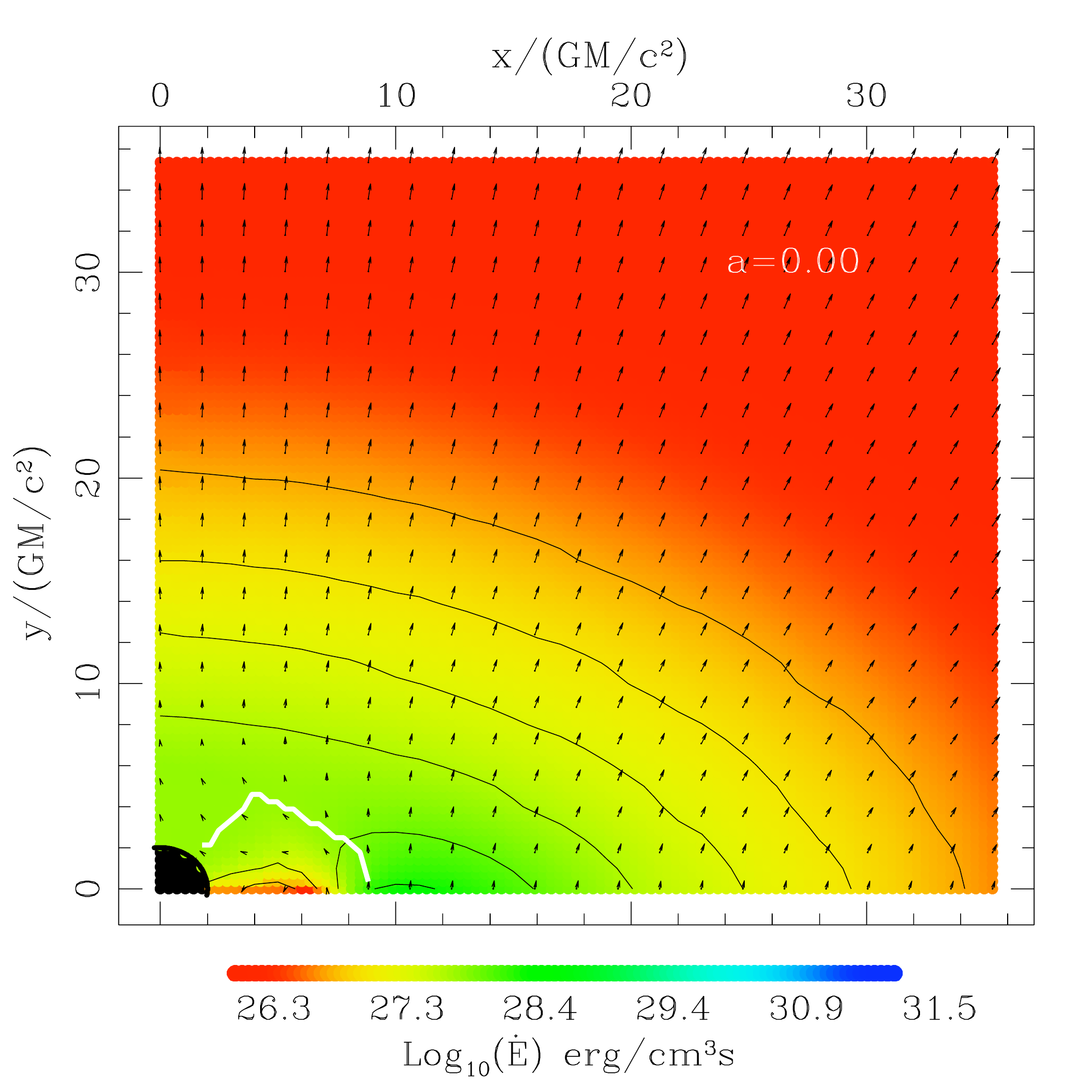} \hfill
\includegraphics[height=0.46\textwidth]{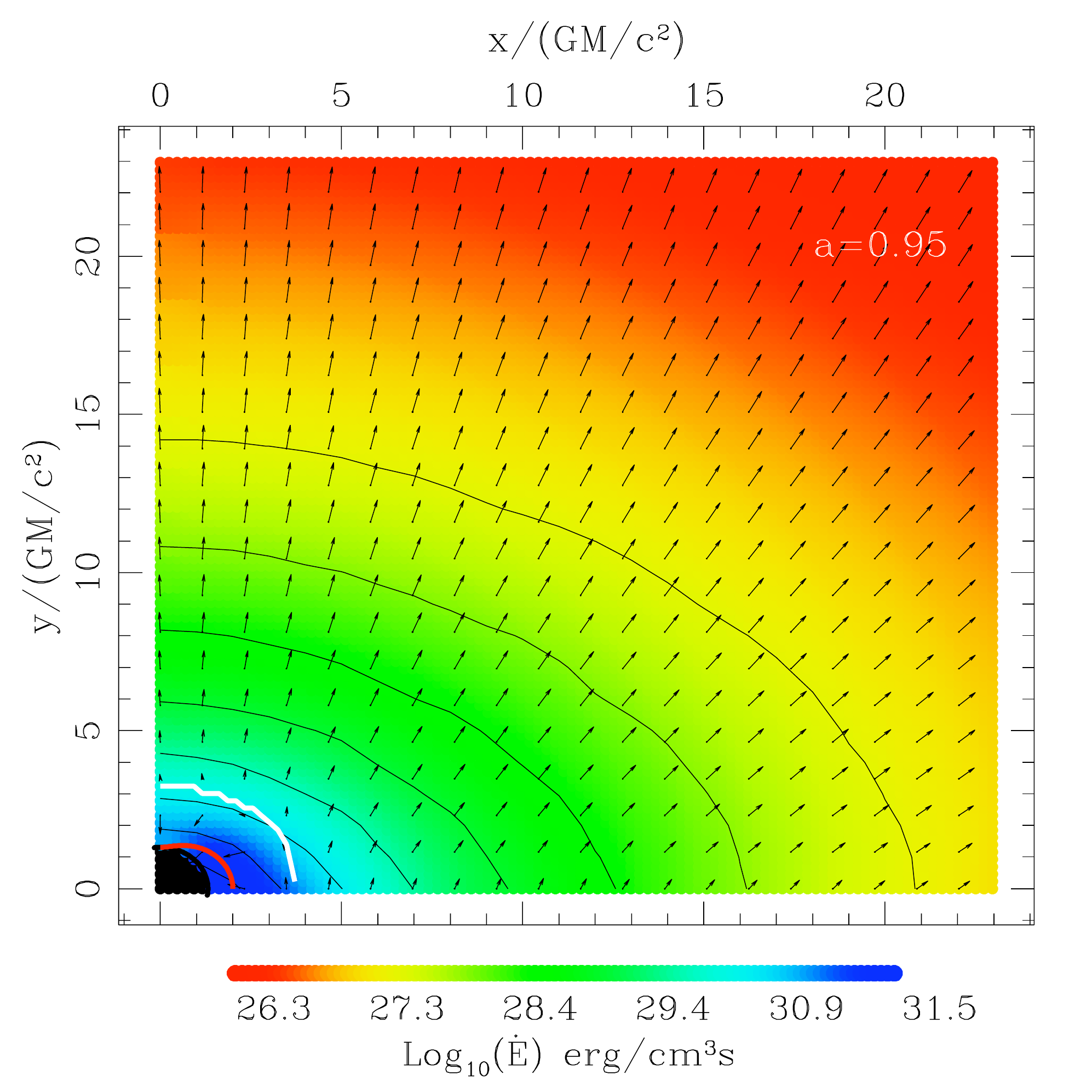}
\caption{Spatial distribution of the energy deposition rate 
by $\nu\bar\nu$ annihilation. 
This example assumes the accretion rate $\dot{M}=1M_{\odot}$s$^{-1}$,
the disk viscosity parameter $\alpha=0.1$, and the black hole
mass $M=3M_{\odot}$; Model A is assumed for the neutrino spectrum (see
the text).
\emph{Left panel:} $a=0$. \emph{Right panel}: 
$a=0.95$ ($a=1$ corresponds to the maximally rotating black hole). 
Note that the energy deposition rate is much higher in the case of $a=0.95$.
The arrows show the specific momentum of 
the $e^\pm$ plasma injected by $\nu\bar\nu$ annihilation.
The white curve is where the radial component of the injected momentum 
changes sign. This boundary gives an idea of the region where the 
injected plasma is lost into the black hole.
}
\end{figure}

\begin{figure}[tp]
\includegraphics[height=0.46\textwidth]{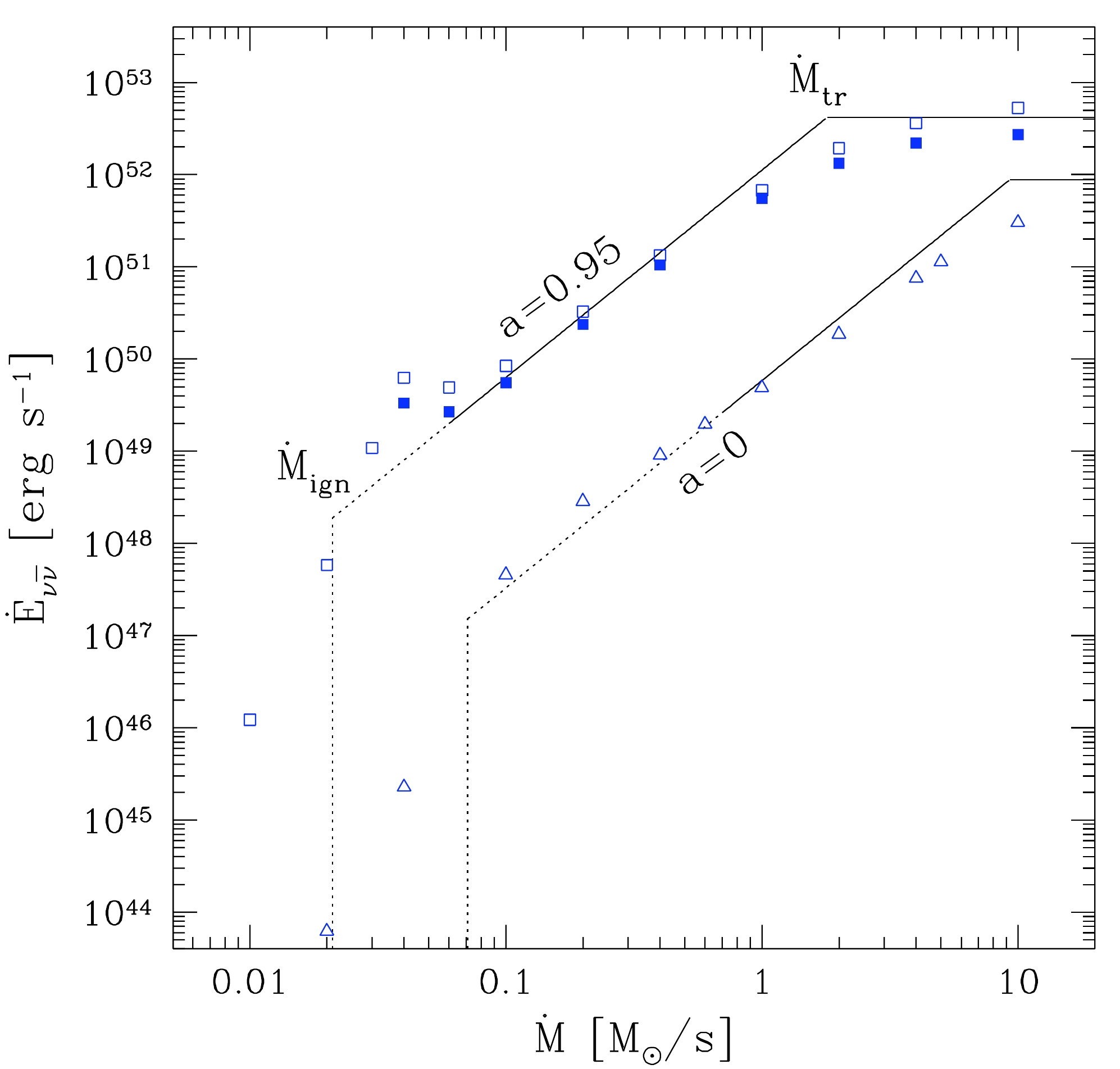} \hfill
\includegraphics[height=0.46\textwidth]{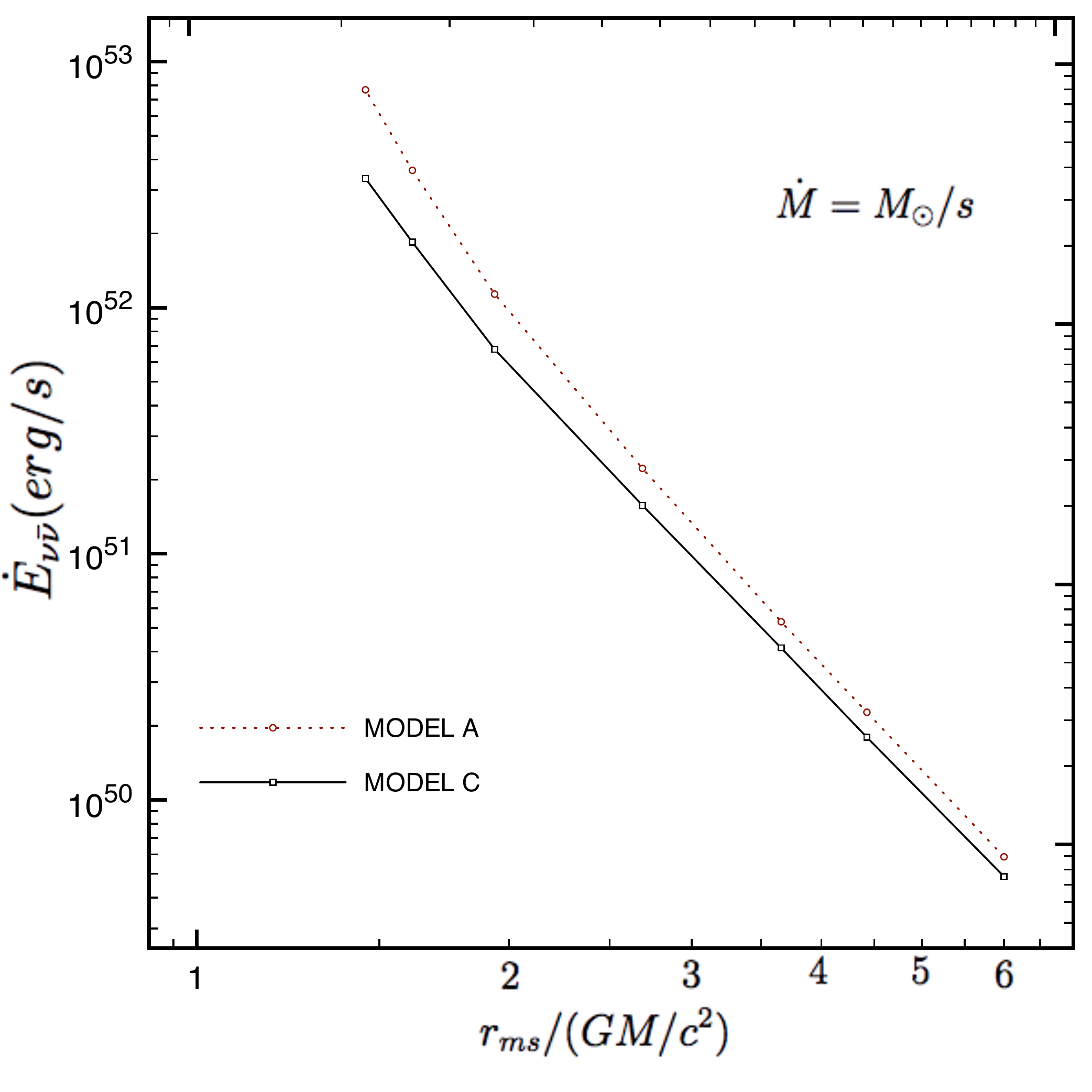}
\caption{ \emph{Left panel:} Total energy deposition rate due to $\nu\bar\nu$ annihilation
outside the black-hole horizon, $\dE$, as a function
of the disk accretion rate, $\dM$. The two characteristic accretion rates
$\dMign$ and $\dMtr$ depend on the viscosity parameter 
$\alpha$
(see eq.~\ref{eq:pw}); $\alpha=0.1$ is chosen in this figure.
The black hole is assumed to have mass $M=3$\,M$_\odot$.
$\dE$ strongly depends on the spin parameter of the black hole;
the results are shown for two cases:
$a=0$ (triangles) and $a=0.95$ (squares). The uncertainty in the vertical
structure of the accretion disk leads to a small uncertainty in $\dE$
as illustrated by two extreme models: Model~A (open symbols) and Model~B
(filled symbols), see the text for details. The results of both models
are well approximated by Model~C 
(eq.~\ref{eq:Edot}), which is shown in the figure by line; the line is
dotted at low $\dM$ where the disk is transparent to neutrinos.
\emph{Right panel}:  Dependence of the energy deposition rate  $\dE$
on the black hole spin for a fixed $\dot M=1M_{\odot}/s$. Instead of using 
the spin parameter $a$ directly, it is 
more instructive
to plot $\dE$ versus the radius of the last (marginally stable) orbit $\rms$. 
Then one can see the power-law dependence of $\dE$ on $\rms$: 
$\dE \varpropto \rms^{-4.7}$. 
The radius $\rms$ is a function of $a$, 
for non-rotating black holes 
$\rms=6 GM/c^{2}$ and for maximally rotating black holes $\rms=GM/c^{2}$.
}
\end{figure}

\section{Results}
Figure~1 shows two examples of the 
spatial distribution of the energy deposition rate 
by $\nu\bar\nu$ annihilation. Integration of this distribution 
over volume outside the black hole
gives the total energy deposition rate $\dot{E}_{\nu\bar\nu}$.
We performed this calculation for various $\dot{M}$ (Fig. 2).
Our results show that $\dE$
 is well approximated by a simple formula,
\begin{equation}
\label{eq:Edot}
  \dE\approx 9\times 10^{51}\,\xms^{-4.7}\,
   \times
   \left\{ \begin{array}{ll}
  0                  &  \dM<\dMign \\
   \dot{m}^{9/4} \;\;
                     &  \dMign<\dM<\dMtr \\
  \dot{m}_{\rm trap}^{9/4}\;\; 
    &  \dM>\dMtr \\
                \end{array}
        \right\}
              \,\frac{\small erg}{\small s}
\end{equation}
where $\dot{m}=\dM/M_{\odot}$s$^{-1}$ and 
$\xms=\rms(a)(2GM/c^2)^{-1}$.
Derivation of the scaling of $\dE$ with $\dot{m}$ and $\xms$ is given
in \cite{Beloborodov:2008}. The dependence of $\dE$ on the black hole spin is huge: 
$\xms^{-4.7}$ varies by a factor of 170 for $0<a<0.95$.
Note that $\alpha$ (viscosity parameter of the disk) enters the result 
only through $\dMign$ and $\dMtr$ (eq.~\ref{eq:pw}).

The efficiency of $\nu\bar\nu$ annihilation can be defined as 
$\epsilon=\dE/L$ where $L$ is the total neutrino luminosity of the disk.
For example $a=0.95$ (which corresponds to $\xms\approx 1$)
gives~$L\approx 0.1\dM c^2$~and \begin{equation}
  \epsilon\approx 0.05\left(\frac{\dM}{M_\odot{\rm ~s}^{-1}}\right)^{5/4},
  \qquad \dMign<\dM <\dMtr.
\end{equation}
The observed GRB luminosity $\Lobs$ can be supplied by 
$\nu\bar\nu$ annihilation around 
a rapidly spinning black hole ($a=0.95$) if 
$\dM>0.38 (M_\odot/s)(\Lobs/10^{51}{\rm ~erg/s})$, which is in the 
range of plausible accretion rates in GRB central engines.

Finally, note that $\dE$ is defined as the \emph{total} energy deposition 
rate outside the event horizon.
A fraction of the created $e^\pm$ plasma falls
into the black hole
and not contribute to the observed explosion (Figure 1).
The corresponding refinement of $\epsilon$
depends on the plasma dynamics outside the disk, which is affected by 
magnetic fields and is hard to calculate without additional assumptions.

\bibliographystyle{aipproc}   
\bibliography{huntsville_proc}

\end{document}